\documentclass[twocolumn,letter]{jpsj2}

\usepackage{graphicx}
\usepackage{dcolumn}
\usepackage{bm}
\def\i{{\rm i}}
\def\d{{\rm d}}
\def\e{{\rm e}}
\def\vector#1{{\bf #1}}
\def\vq{{\vector q}}
\def\vk{{\vector k}}
\def\vr{{\vector r}}

\def\ve{{\vector e}}
\def\dps{\displaystyle}

\def\hightc{{high-$T_{\rm c}$ }}

\def\parallelsl{{/\hspace{-0.7ex}/}}

\title{
Phase Fluctuations and Kosterlitz-Thouless Transition in \\
Two-Dimensional Fulde-Ferrell-Larkin-Ovchinnikov \\
Superconductors 
}

\author{Hiroshi {\sc Shimahara}}

\inst{
Faculty of Science, Hiroshima University, 
Higashi-Hiroshima 739-8530
\footnote{Present address: Department of Quantum Matter Science, 
ADSM, Hiroshima University, Higashi-Hiroshima 739.} 
}

\recdate{February 12, 1998}

\abst{
Effect of the phase fluctuations of the order parameter 
on the stability of the Fulde-Ferrell-Larkin-Ovchinnikov (FFLO) states 
are examined in exactly two-dimensional (2D) type-II superconductors 
with cylindrically symmetric Fermi surface 
on the basis of a generalized Ginzburg-Landau theory. 
It is found that 
for the FFLO states with oscillations in a single direction, 
not only the long-range order but also quasi-long-range order (QLRO), 
which is characterized by a power law decay of the order parameter 
correlation function, 
is suppressed by the phase fluctuations at any finite temperatures. 
On the other hand, for the FFLO states with order parameter structures 
such as triangular and square lattices, 
it is shown that the QLRO is possible as the uniform BCS state. 
Systems with anisotropy in the Fermi surface and pairing are also 
discussed. 
}

\kword{
Fulde-Ferrell-Larkin-Ovchinnikov state, 
nonuniform superconductivity,
organic superconductor, 
Kosterlitz-Thouless transition, 
phase fluctuations, 
Pauli paramagnetic limit, 
Chandraseker and Clogston limit, 
quasi-long-range order
}

\begin{document}
\sloppy
\maketitle

The nonuniform superconducting state called 
the Fulde-Ferrell-Larkin-Ovchinnikov (FFLO) state 
is characterized by spatial oscillations of the order parameter 
due to the Zeeman energy 
additional to the spatial dependence due to the vortices 
around the magnetic flux lines. 
It is known that the FFLO state is stable at high magnetic fields 
in type-II superconductors within a mean field theory 
if some ideal conditions are satisfied.~\cite{ful64,lar64} 
However, there has not been any theory which has examined this state beyond 
the mean field approximation, although the two-dimensional (2D) systems are 
important when one considers the FFLO state, as we shall see below. 
In this paper, we examine the effect of the phase fluctuations in the 2D 
systems on the stability of the FFLO state.

For the FFLO state to occur in a type-II superconductor, 
the following two conditions need to be satisfied. 
(1) Orbital pair breaking effect is sufficiently weakened by some 
mechanism, so that the Zeeman energy reaches a value of the order of 
the zero field gap, near the upper critical field.~\cite{gru66} 
(2) The sample is in a clean limit.~\cite{sar69,tak69} 
In the alloy type-II superconductors, these conditions are 
difficult to be satisfied, 
but it is possible to satisfy them in the organic and \hightc oxide 
superconductors 
because of the strong electron correlations, narrow electron bands, and 
quasi-low-dimensionality. 
In low-dimensional superconductors, 
the orbital pair breaking effect is weakened by applying 
the field in any direction parallel to the most conducting 
layer,~\cite{bul73,aoi74,buz83,leb86,dup93,bur94,shi94,dup95,mak96,buz96a,buz97,shi97a,shi97b,shi98a} 
with a sufficient accuracy.~\cite{shi97b} 

Further, the enhancement of the FFLO critical field due to 
the anisotropies of the Fermi surface (including the low-dimensional 
nature) and 
the pairing~\cite{bul73,aoi74,buz83,leb86,dup93,bur94,shi94,dup95,mak96,buz96a,buz97,shi97a,shi97b,shi98a,mat94} 
would increase 
the possibility of the FFLO state being found in the organic and oxide 
superconductors as discussed in detail in our previous 
papers.~\cite{shi94,shi97a} 
This is not merely a quantitative issue, 
but makes a qualitative difference in the realization of the FFLO state, 
if one takes into account negative effects on the FFLO state, 
such as internal field enhancement.~\cite{bur94,shi94,shi97a}

Therefore, low-dimensional exotic type-II superconductors, 
such as the organic and copper oxide superconductors, 
might be good candidates of the FFLO superconductor. 
At present, we do not have any experimental evidence of the FFLO state, 
but some unusual experimental facts in organic superconductors 
can be explained as a consequence 
of the FFLO state,~\cite{shi98a,lee97,mur85,mur88,lyb90,yag84} 
although other explanations such as the triplet pairing superconductivity 
are also possible.~\cite{gor86,bur87,she91,bel97}

In this paper, we regard the quasi-one-dimensional organic superconductors 
such as ${\rm (TMTSF)_2X}$ and ${\rm (DMET)_2X}$ 
as quasi-2D systems, in the sense that 
(1) Fermi surfaces are sufficiently deformed by the application of pressure 
and so on, so that the nesting instabilities are suppressed, and 
(2) there are no flat portions in their Fermi surfaces, 
which would cause the FFLO critical field at $T=0$ to diverge.

In many organic superconductors, 
especially in those with low zero-field transition temperatures 
to the superconductivity, 
interlayer electron transfers are strong enough to justify the mean field 
treatment at low temperatures of interest. 
In the theories of the FFLO state in the low-dimensional 
systems,~\cite{bul73,aoi74,buz83,leb86,dup93,bur94,shi94,dup95,mak96,buz96a,buz97,shi97a,shi97b,shi98a} 
such interlayer couplings are assumed implicitly. 
However, in some organic superconductors and the \hightc oxide 
superconductors, the interlayer coupling is not so 
strong.~\cite{ish97,osh88,bro95,tan97} 
For example, $\kappa$-${\rm (BEDT}$-${\rm TTF)_4Hg_{2.89}Br_8}$ has 
very small interlayer coupling. 
The observed critical field reaches about 5 times the Pauli paramagnetic 
limit (Chandraseker and Clogston limit) 
estimated from the zero-field transition temperature using a simplified 
formula.~\cite{lyb90} 
The estimated value should be modified by many effects, 
but it seems difficult to explain such an extremely large value 
within the mean field theory if a singlet pairing is assumed. 
As a possibility, 
low-dimensional fluctuations may play an important role in the large ratio 
of the upper critical field to the zero field transition temperature.

In such low-dimensional superconductors, it is known that the phase 
fluctuations of long wavelength would suppress the long-range order 
(LRO) of the superconductivity at any finite temperature, 
but the quasi-long-range order (QLRO) may occur below a finite 
Kosterlitz-Thouless (KT) transition temperature.~\cite{kos73,kos74} 
Here, the QLRO is characterized by a power law decay of the order parameter 
correlation function for long distances. 
When small interlayer transfer is introduced, the LRO of the superconductivity 
occurs at the temperatures below a finite transition temperature 
close to the KT transition temperature, 
which would be much smaller than the mean field value of the transition 
temperature. 
Also for the FFLO state, the phase fluctuation must have a serious effect 
on the LRO in the low-dimensional systems. 
In two dimensions, 
it is clear that the LRO of the FFLO superconductivity does not occur 
at finite temperatures, because of the existence of continuous symmetry, 
as in the BCS superconductivity. 
However, it is not yet clear whether the QLRO of the FFLO superconductivity 
occurs, since the nonuniformity of the FFLO state may affect 
the phase fluctuations.

We first note that the FFLO state could take various structures 
other than simple oscillations in a single direction at high magnetic 
fields.~\cite{shi98a}
The most general form of the order parameter $\psi(\vr)$ is written as 
\def\eqpsi
{(1)}
$$
     \psi(\vr) = \sum_{m} \psi_0^{(m)} \e^{\i \vq_m \cdot \vr} 
     \eqno\eqpsi
     $$
near the FFLO critical field, 
where $\vq_m$'s are some of the optimum wave vectors which yeild 
the highest critical field. 
For the cylindrically symmetric Fermi surface, 
we have an infinite number of optimum vectors for $s$-wave pairing, 
while we have four such vectors for $d$-wave pairing. 
We write their magnitude as $q = |\vq_m|$, 
because they are the same. 
In this paper, since we consider only periodic solutions, 
only a finite number of $\vq_m$'s are considered in eq.~{\eqpsi}. 
The FFLO state with spatial oscillations in a single direction is 
expressed by eq.~{\eqpsi} in which only a single wave vector $\vq_1$ is 
considered, or two antiparallel wave vectors ($\vq_1 = - \vq_2$) 
are taken into account. 
We call such structures one-dimensional (1D) structures. 
On the other hand, when we consider more than one $\vq_m$ in eq.~{\eqpsi} 
which includes two $\vq_m$'s neither parallel nor antiparallel to each 
other, we call it a 2D structure. 
In the FFLO state, the spatial symmetry is broken spontaneously 
on the appearance of the order parameter.

In the mean field approximation, 
the optimum form among those expressed by eq.~{\eqpsi} is determined 
by the terms of the fourth order of $\psi(\vr)$ in the free 
energy.~\cite{lar64,shi98a} 
It has usually been believed that a 1D structure is optimum, 
since Larkin and Ovchinnikov found that it is optimum in a spherically 
symmetric system,~\cite{lar64} 
but in practice, it may not be the case in other systems. 
For example, in a cylindrically symmetric system, 
the triangular, square, and hexagonal states occur as the temperature 
decreases for the $s$-wave pairing, 
while the square states occur at low temperatures 
for the $d$-wave pairing.~\cite{shi98a} 
This is not merely due to a geometrical effect, 
but due to the high FFLO critical field in low-dimensions, 
for which the 2D structures are favored over the 1D structures 
due to spin-polarization energy.

In general, the optimum structure may depend on 
the symmetries and other details of the Fermi surface and the pairing, 
and the strength of the field. 
Thus, we consider the general form of eq.~{\eqpsi} as the stable 
structure of the order parameter in the absence of the phase fluctuations, 
assuming that it is known a priori.

We first construct a phase Hamiltonian from a generalized form of 
the Ginzburg-Landau (GL) Hamiltonian. 
We assume the rotational symmetry of the system and 
define the Hamiltonian as 
\renewcommand{\arraystretch}{1.5}
\def\eqGLH
{(2)}
$$
     H[\psi]= \int \! \! \d^2 r ~ [
                      \frac{1}{2} c |\nabla \psi(\vr)|^2 
                    + \frac{1}{2} d |\Delta \psi(\vr)|^2 ] , 
     \eqno\eqGLH
     $$
where we have taken into consideration only second order terms 
related to the phase fluctuations. 
The second term has been added so that the energy is bounded when $c < 0$. 
This is the most compact form which allows nonuniformity of the order 
parameter in the absence of the vector potential. 
The detailed form of the GL free energy near the tri-critical point 
was microscopically calculated by Buzdin and KachKachi,~\cite{buz97} 
but we take a more phenomenological viewpoint. 
Substituting eq.~{\eqpsi} into eq.~{\eqGLH}, 
we obtain an expression for the energy as a function of $q$. 
When $c<0$, a finite value of $ q = \sqrt{|c|/(2d)}$ 
gives the energy minimum, and the nonuniform state is stabilized.

For finite temperatures, 
we introduce the phase fluctuation 
in the order parameter $\psi(\vr)$ by 
\def\eqpsiwithphase
{(3)}
$$
     \psi(\vr) = {\bar \psi}(\vr) \e^{\i \phi(\vr)} , 
     \eqno\eqpsiwithphase
     $$
where ${\bar \psi(\vr)}$ is the order parameter at $T = 0$ given by 
eq.~{\eqpsi} with the above value of $q$, 
and $\phi(\vr)$ is a real function expressing the phase fluctuations. 
We neglect fluctuations in the magnitude of the order parameter at 
sufficiently low temperatures. 
Assuming that $\phi(\vr)$ is small and varies slowly over a distance 
scale much larger than the inverse of $q$, 
the phase Hamiltonian is derived as 
\def\eqphaseH
{(4)}
$$
     H_{\phi} = \frac{1}{2} d 
     \sum_{m} |\psi_0^{(m)}|^2 \int \! \! \d^2 r 
     [(\Delta \phi)^2 + 4 (\vq_m \cdot \nabla \phi)^2 ] . 
     \eqno\eqphaseH
     $$
When all values of $\psi_0^{(m)}$ are equal to $\psi_0$, 
\def\eqphaseH
{(5)}
$$
     H_{\phi} = C_{\phi} \int \! \! \d^2 r 
     [(\Delta \phi)^2 
     + \frac{4}{N_q} \sum_m (\vq_m \cdot \nabla \phi)^2 ] , 
     \eqno\eqphaseH
     $$
where $N_q$ denotes the number of $m$'s which are taken in the summation 
in eq.~{\eqpsi}, and $C_{\phi} \equiv \frac{1}{2} d \cdot N_q |\psi_0|^2$. 
Here, we find a remarkable difference in the second term 
between the 1D and 2D structures: 
for the 1D structures, the second term is proportional to 
$(\vq \cdot \nabla \phi)^2$, while for the 2D structures, 
it is proportional to $q^2 (\nabla \phi)^2$.

We should note that the FFLO states have many line nodes in the real space, 
which divide the real space into many sections or cells. 
Since the phase $\phi$ is not actually defined where the amplitude vanishes, 
it appears that the phase can be discontinuous there. 
However, if the phase jumps between the sections divided by the line nodes, 
Josephson current is induced and causes the phase difference to vanish rapidly. 
If we regard the connection between the sections as a tunnel junction, 
it is almost transparent to electron tunnelling, 
because no barrier actually exists between the sections. 
Thus, the phase difference must be negligibly small even if it exists. 
Therefore, it is reasonable to assume 
that the phase function $\phi$ is a smooth continuous function over 
the sections.

The spatial correlation of the order parameter is defined by 
\def\eqcorr
{(6)}
$$
     C(\vr) = <\!\![{\tilde \psi}(\vr)]^{*} {\tilde \psi}(0) \!\!> 
     \eqno\eqcorr
     $$
with ${\tilde \psi}(\vr) \equiv \psi_0 \e^{\i \phi(\vr)}$. 
Since eq.~{\eqphaseH} is bilinear in $\phi(\vr)$, we have 
\renewcommand{\arraystretch}{2}
\def\eqcorrphi
{(7)}
$$
     \begin{array}{rcl} 
     C(\vr) \!\! & \!\! = \!\! & \!\! 
     |\psi_0|^2 \exp[\dps{-\frac{1}{2}<\!\![\phi(\vr)-\phi(0)]^2\!\!>}] \\
     \!\! &  \!\! = \!\! & \!\! 
     |\psi_0|^2 \exp[\dps{ -\frac{1}{L^4} \sum_{\vk} 
     |\e^{\i \vk \cdot \vr} - 1|^2 <\!\!|\phi_\vk|^2\!\!>  }] , 
     \end{array}
     \eqno\eqcorrphi
     $$
where $L$ is the linear dimension of the system. 
The average $<\!\!|\phi_\vk|^2\!\!>$ is calculated from eq.~{\eqphaseH} as 
\def\eqavephik
{(8)}
$$
     <\!\!|\phi_\vk|^2\!\!> = 
     \frac{L^2 T}{C_{\phi}} \frac{1}
     {\dps{k^4 + \frac{4}{N_q} \sum_m (\vq_m \cdot \vk)^2}} . 
     \eqno\eqavephik
     $$

The states with the 1D structures are given by 
$N_q = 1$ and $\vq_1 = \vq$, or $N_q = 2$ and $\vq_1 = -\vq_2 = \vq$. 
For the both states, we find 
\def\eqavephikoneD
{(9)}
$$
     <\!\!|\phi_\vk|^2\!\!> = 
     \frac{L^2 T}{C_{\phi}} \frac{1}
     {\dps{k^4 + 4 (\vq \cdot \vk)^2}} . 
     \eqno\eqavephikoneD
     $$
We substitute this expression into the integral in eq.~{\eqcorrphi}. 
Unless $\vk \perp \vq$, the integrand is proportional to $1/k^2$ for 
small $k$, and then the integral over $k$ diverges logarithmically. 
However, in practice, we have a stronger divergence from the 
contribution near $\vk \perp \vq$. 
Since the essentially important contribution comes from 
the vicinity of $\vk \perp \vq$, 
we put $\varphi = \pi/2$ in the factor $|\e^{\i \vk \cdot \vr} - 1|^2$, 
where $\varphi$ is the angle between $\vk$ and $\vq$, 
except when the angle $\theta$ between $\vq$ and $\vr$ is not very small. 
Then, this factor introduces the cutoff momentum of the order of 
$1/(r\sin\theta)$ for small $k$. 
For long distances such as $r \gg \xi \sim 1/q$, 
the factor $|\e^{\i \vk \cdot \vr} - 1|^2$ can be replaced by 
the average value of 2. 
Thus, we obtain the short-range order 
\def\eqSRO
{(10)}
$$
     C(\vr) = |\psi_0|^2 \exp(-r/{\tilde \xi}) 
     \eqno\eqSRO
     $$
with $ {\tilde \xi} \equiv 2\pi q C_{\phi}/(T\sin\theta)$. 
When $\vr \parallelsl \vq$, since the integrand vanishes 
at $\varphi = \pi/2$ because of the factor $|\e^{\i \vk \cdot \vr} - 1|^2$, 
we obtain a logarithmic divergence of the integral in the long 
distance limit and the power law decay of the correlation function. 
Except for this direction $\vr \parallelsl \vq$, we only have a short-range 
order for the FFLO states with the 1D structures.

On the other hand, for the states with 2D structures 
such as the triangular lattices and the square lattice,~\cite{shi98a} 
a QLRO occurs in two dimensions. 
In the denominator of the integrand of eq.~{\eqavephik}, 
the terms in the summation $\sum_m (\vq_m \cdot \vk)^2$ 
do not simultaneously vanish when $k \ne 0$. 
Hence, for long wavelengths, the first term in the denominator 
proportional to $k^4$ can be omitted, and we have a logarithmic divergence 
of the integral in eq.~{\eqcorrphi}, 
which leads to the power law decay of the correlation function 
in the long distance limit. 
There are three possible 2D structures, that is, the triangular state: 
$N_q = 3$, $\vq_1 = q(1,0)$, $\vq_2 = q(-1/2,\sqrt{3}/2)$, 
$\vq_3 = q(-1/2,-\sqrt{3}/2)$, 
the square state: 
$N_q = 4$, $\vq_1 = q(1,0)$, $\vq_2 = q(0,1)$, $\vq_3 = -\vq_1$, 
$\vq_4 = -\vq_2$, 
and the hexagonal state: 
$N_q = 6$, $\vq_1 = q(1,0)$, $\vq_2 = q(-1/2,\sqrt{3}/2)$, 
$\vq_3 = q(-1/2,-\sqrt{3}/2)$, $\vq_4 = -\vq_1$, $\vq_5 = -\vq_2$, 
$\vq_6 = -\vq_3$. 
For these states the Hamiltonian is reduced to 
\def\eqphaseHtwoD
{(11)}
$$
     H_{\phi} = \frac{1}{2} K_{\phi} \int \! \! \d^2 r 
     (\nabla \phi)^2 
     \eqno\eqphaseHtwoD
     $$
with $K_{\phi} = 2 q^2 C_{\phi}$. 
The summation $\sum_{\vk}$ in eq.~{\eqcorrphi} is taken 
between cutoff momenta $k \sim 1/r$ and $k \sim 1/a$, 
where $a$ is a length of the order of the lattice constant. 
The former cutoff is due to the factor $|\e^{\i \vk \cdot \vr} - 1|^2$. 
Therefore, we obtain a power law decay of the correlation function 
\def\eqCpowerlaw
{(12)}
$$
     C(\vr) = |\psi_0|^2 (\frac{a}{r})^{\eta} 
     \eqno\eqCpowerlaw
     $$
with $\eta = T/(2\pi K_{\phi}) = T/(2\pi |c| N_q |\psi_0|^2)$. 
The KT criterion for the stability of the QLRO against 
free vortex formation is roughly written as 
\def\eqTKT
{(13)}
$$
     T < \frac{\pi}{2} K_{\phi}
     = \frac{\pi}{2} |c| N_q |\psi_0|^2 = T_{\rm KT} . 
     \eqno\eqTKT
     $$
Since the stiffness constant $K_{\phi}$ is reduced by the polarization 
of vortex pairs, the actual value of $T_{\rm KT}$ is smaller than 
that given above. 
In the magnetic field, 
the vortex formation would reduce the energy by an amount proportional 
to the number of the vortices, since they are accompanied by 
the spin and orbital polarizations. 
Thus, the KT transition temperature decreases with the magnetic field, 
and vanishes at a critical magnetic field. 
The phase diagram will be investigated in a future study.

We briefly comment on the phase fluctuations in three dimensions. 
In this case, the spatial correlation of the phase diverges 
logarithmically for the FFLO states with 1D structures. 
Thus, it may be conjectured that the LRO's of the FFLO states with 
1D structures are suppressed by the phase fluctuations 
at finite temperatures even in three dimensions, and a QLRO might occur. 
This subject will be discussed in more detail in another paper.

Lastly, we discuss anisotropic systems. We have studied the isotropic 
Hamiltonian given by eq.{\eqGLH} in two dimensions, 
but it would be interesting to extend our calculation to anisotropic 
systems for more realistic models. 
For example, a model Hamiltonian 
\renewcommand{\arraystretch}{1.5}
\def\eqGLHani
{(14)}
$$
     \begin{array}{rcl}
     H[\psi] \!\! & \!\! = \!\! & \!\! 
                \dps{ \int \! \! \d^2 r ~ [
                      \frac{1}{2} c |\nabla \psi(\vr)|^2 
                    + \frac{1}{2} (d_0 + 3 d_1) |\Delta \psi(\vr)|^2 } \\
             & &    - d_1 \sum_i |(\ve_i \cdot \nabla)^2 \psi(\vr)|^2 ]  
     \end{array}
     \eqno\eqGLHani
     $$
describes a system with tetragonal symmetry, 
where 
$\ve_1 = -\ve_4 = (1,0)$, 
$\ve_2 = -\ve_3 = (0,1)$ and 
$d_0 > d_1 > 0$. 
For an order parameter of the form 
$\psi(\vr) = \psi_0 \exp(\i \vq \cdot \vr)$, 
the energy of the system is given by 
\def\eqEani
{(15)}
$$
     E(q,\varphi)= L^2 |\psi_0|^2 [
                      \frac{1}{2} c q^2 
                    + \frac{1}{2} q^4 
                    \{ d_0 - d_1 \cos(4 \varphi) \} ] , 
     \eqno\eqEani
     $$
which becomes minimum 
at $\varphi = 0, \pi/2, \pi, 3\pi/2$ and 
$$ 
     q = \sqrt{|c|/2(d_0-d_1)} . 
     $$ 
The phase Hamiltonian is derived as 
\def\eqphaseHani
{(16)}
$$
     \begin{array}{rcl}
     \lefteqn{ \dps 
     H_{\phi} = 
     \sum_{m} |\psi_0^{(m)}|^2 \int \! \! \d^2 r \,\,      
     [   4 d_1 q^2 (\nabla \phi)^2                     } \\
     & & 
       + 2 (d_0 - 3 d_1) (\vq_m \cdot \nabla \phi)^2 
     + \frac{1}{2} (d_0 + 3 d_1) (\Delta \phi)^2         \\
     & & 
     - 2 d_1 \{ (\partial_x^2 \phi)^2 + (\partial_y^2 \phi)^2 \} ]
     \end{array}
     \eqno\eqphaseHani
     $$
It is easily verified that 
the QLRO's of the FFLO states with 1D structures are 
stabilized as a local free energy minimum, 
differently from the isotropic model, but 
the stiffness constants are smaller 
than that in the state with a 2D structure, 
i.e., a square state in this case.~\cite{shi98a} 
Therefore, 
the KT transition temperature to the FFLO state with the 1D structures 
is smaller than that to the state with the 2D structure also 
in anisotropic systems.

The results of this paper are explained as follows. 
In 1D structures, the phase fluctuations with small wave vectors $\vk$ 
perpendicular to $\vq$ are regarded as the fluctuations of the direction 
of $\vq$. 
On the other hand, in 2D structures, any phase fluctuations 
are inevitably accompanied by fluctuations of the magnitude of $\vq_m$'s 
in addition to the fluctuations of the directions. 
Thus, the stiffness constants in the 2D structures are usually larger 
than those in the 1D structures, and the phase fluctuations affect 
the FFLO states with the 1D structures more seriously.

In conclusion, we have examined the effect of the phase fluctuations 
on the FFLO state in exactly 2D type-II superconductors. 
From the generalized GL Hamiltonian in which a nonuniform state is 
stabilized at $T = 0$ even in the absence of the vector potential, 
we have derived an effective Hamiltonian for the slowly varying 
phase of the order parameter. 
On the basis of the phase Hamiltonian, we have obtained the following 
results. 
(1) When the FFLO state at $T = 0$ has an order parameter which oscillates 
in a single direction, such as $\Delta \sim \cos(\vq \cdot \vr)$ and 
$\Delta \sim \exp(\i \vq \cdot \vr)$, 
the LRO and even the QLRO's are unstable at finite temperatures 
because of the phase fluctuations in the directions perpendicular to $\vq$. 
(2) When the FFLO state at $T = 0$ has an order parameter with a 2D structure 
such as the triangular and square lattices, since the phase Hamiltonian has 
the same form as in the case of uniform superconductivity, 
a QLRO occurs below a finite KT transition temperature, 
although the LRO is suppressed. 
Therefore, in 2D FFLO superconductors with very weak interlayer coupling 
and cylindrically symmetric Fermi surface, 
the structure of the order parameter must be that of the 2D lattices, 
and not simple oscillations in a single direction. 

\vspace{\baselineskip}

\section*{Acknowledgements}

The author would like to thank Professor T. Ishiguro for useful information 
on experimental facts in low-dimensional organic superconductors.



\begin{thebibliography}{99}
\bibitem{ful64}
  P. Fulde and R. A. Ferrell: Phys. Rev. {\bf 135} (1964) A550. 
\bibitem{lar64}
  A. I. Larkin and Yu. N. Ovchinnikov: 
  Zh. Eksp. Teor. Fiz. {\bf 47} (1964) 1136, 
  translation: Sov. Phys. JETP, {\bf 20} (1965) 762. 
\bibitem{gru66}
  L. W. Gruenberg and L. Gunther: Phys. Rev. Lett. 
  {\bf 16} (1966) 996. 
\bibitem{sar69}
  G. Sarma and D. Saint-James: {\it Communication to the Conf. 
  on the Phys. of Type II Superconductivity, }
  Western Reserve University, Cleveland, (Ohio); 
  D. Saint-James, G. Sarma, and E. J. Thomas: 
  {\it Type II Superconductivity} (Pergamon Press, Oxford, 1969). 
\bibitem{tak69}
  S. Takada and T. Izuyama: Prog. Theor. Phys. {\bf 41} (1969) 635; 
  S. Takada: Prog. Theor. Phys. {\bf 43} (1970) 27. 
\bibitem{bul73}
  L. N. Bulaevskii: Zh. Eksp. Teor. Fiz. {\bf 65} (1973) 1278, 
  translation: Sov. Phys. JETP {\bf 38} (1974) 634. 
\bibitem{aoi74}
  K. Aoi, W. Dieterich and P. Fulde: Z. Phy. {\bf 267} (1974) 223. 
\bibitem{buz83}
  A. I. Buzdin and V. V. Tugushev: Zh. Eksp. Teor. Fiz. {\bf 85}, 
  735 (1983), translation: Sov.Phys. JETP {\bf 58} (1983) 428. 
\bibitem{leb86}
  A. G. Lebed': Pis'ma Zh. Eksp. Teor. Fiz. {\bf 44} (1986) 89, 
  translation: Sov. Phys. JETP Lett. {\bf 44} (1986) 144. 
\bibitem{dup93}
  N. Dupuis, G. Montanbaux and C. A. Ra. S${\acute {\rm a}}$ de Melo: 
  Phys. Rev. Lett. {\bf 70} (1993) 2613. 
\bibitem{bur94}
  H. Burkhardt and D. Rainer: Ann. Physik {\bf 3} (1994) 181. 
\bibitem{shi94}
  H. Shimahara: Phys. Rev. B {\bf 50} (1994) 12760. 
\bibitem{dup95}
  N. Dupuis: Phys. Rev. B {\bf 51} (1995) 9074. 
\bibitem{mak96}
  K. Maki and H. Won: Czechoslovak J. Phys. {\bf 46}, 
  Suppl. S2 (1996) 1035. 
\bibitem{buz96a}
  A. I. Buzdin and J. P. Brison: Europhys. Lett. {\bf 35} (1996) 707. 
\bibitem{buz97}
  A. I. Buzdin and H. Kachkachi: 
  Phys. Lett. A {\bf 225} (1997) 341. 
\bibitem{shi97a}
  H. Shimahara: J. Phys. Soc. Jpn. {\bf 66} (1997) 541. 
\bibitem{shi97b}
  H. Shimahara and D. Rainer: J. Phys. Soc. Jpn. {\bf 66} (1997) 3591. 
\bibitem{shi98a}
  H. Shimahara: J. Phys. Soc. Jpn. {\bf 67} (1998) 736. 
\bibitem{mat94}
  S. Matsuo, H. Shimahara and K. Nagai: 
  J. Phys. Soc. Jpn. {\bf 63} (1994) 2499; 
  {\it ibid}: 
  J. Phys. Soc. Jpn. {\bf 64} (1995) 371; 
  H. Shimahara, S. Matsuo and K. Nagai: 
  Phys. Rev. B {\bf 53} (1996) 12284. 
\bibitem{lee97}
  I. J. Lee, M. J. Naughton, G. M. Danner and P. M. Chaikin: 
  Phys. Rev. Lett. {\bf 78} (1997) 3555. 
\bibitem{mur85}
  K. Murata {\it et al.}: 
  Physica B\&C {\bf 135BC} (1985) 515. 
\bibitem{mur88}
  K. Murata {\it et al.}: 
  Synth. Met. {\bf 27} (1988) 341. 
\bibitem{lyb90}
  R. N. Lyubovskaya {\it et al.}: 
  Pis'ma Zh. Tekh. Fiz. (USSR),{\bf 16} (1990) 80. 
\bibitem{yag84}
  E. B. Yagubskii {\it et al.}: 
  Pis'ma v Zh. Eksp. Teor. Fiz. (USSR), {\bf 39} (1984) 275. 
\bibitem{gor86}
  L. P. Gor'kov: 
  Pis'ma v Zh. Eksp. Teor. Fiz. {\bf 44} (1986) 537. 
\bibitem{bur87}
  L. I. Burlachkov, L. P. Gor'kov and A. G. Lebed': 
  Europhys. Lett. {\bf 4} (1987) 941. 
\bibitem{she91}
  I. F. Shegolev and E. B. Yagubskii: 
  Physica C, {\bf 185-189} PT.1, (1991) 360. 
\bibitem{bel97}
  S. Belin and K Behnia: Phys. Rev. Lett. {\bf 79} (1997) 2125. 
\bibitem{ish97}
  T. Ishiguro, H. Ito, Y. Yamauchi, E. Ohmichi, M. Kubota, H. Yamochi,
  G. Saito, M. V. Kartsovnik, M. A. Tanatar, Yu. V. Sushko and 
  G. Yu. Logvenov: 
  Synth. Met. {\bf 85} (1997) 1471. 
\bibitem{osh88}
  K. Oshima, H. Urayama, H. Yamochi and G. Saito: 
  Physica C {\bf 153-155}, PT 2 (1988) 1148. 
\bibitem{bro95}
  J. S. Brooks, S. Uji, H. Aoki, T. Terashima, M. Tokumoto, 
  N. Kinoshita, Y. Tanaka and H. Anzai: 
  Synth. Met. {\bf 70} (1995) 839. 
\bibitem{tan97}
  H. Taniguchi, Y. Nakazawa and K. Kanoda: 
  Synth. Met. {\bf 85} (1997) 1553. 
\bibitem{kos73}
  J. M. Kosterlitz and D. J. Thouless: J. Phys. C {\bf 6} (1973) 1181. 
\bibitem{kos74}
  J. M. Kosterlitz: J. Phys. C {\bf 7} (1974) 1046. 
\end{thebibliography}
\end{document}